\documentstyle[epsf,12pt]{article}
\evensidemargin\oddsidemargin
\newcommand{\alt}{\mathrel{\raisebox{-.6ex}{$\stackrel{\textstyle<}{\sim}$}}}
\newcommand{\agt}{\mathrel{\raisebox{-.6ex}{$\stackrel{\textstyle>}{\sim}$}}}

\def\overlay#1#2{\ifmmode \setbox 0=\hbox {$#1$}\setbox 1=\hbox to\wd 0{\hss
$#2$\hss }\else \setbox 0=\hbox {#1}\setbox 1=\hbox to\wd 0{\hss #2\hss }\fi
#1\hskip -\wd 0\box 1}

\def\nv#1 {\noalign{\vskip#1pt}}

\def\abstract#1{\begin{center}\large\bf Abstract\end{center}
{\narrower\small #1\par}}

\def\gev{{\rm\,GeV}}

\begin{document}
\vspace*{-.6in}
\hfill\vbox{\hbox{\bf MAD/PH/760}
            \hbox{May 1993}}\par

\begin{center}
{\large\bf Phenomenological Implications of the\boldmath{$m_t$} RGE Fixed Point
\\[.2in]
for SUSY Higgs
Boson Searches%
\footnote{Talk presented by V.~Barger at Hawaii LCWS (April, 1993).}}\\[.4in]
{ V.~Barger\rlap,$^{\,a}$ M.~S.~Berger\rlap,$^{\,a}$ P.~Ohmann\rlap,$^{\,a}$
and R.J.N.~Phillips${\,^b}$}\\[.2in]
\it
$^a$Physics Department, University of Wisconsin, Madison, WI 53706, USA\\
$^b$Rutherford Appleton Laboratory, Chilton, Didcot, Oxon OX11 0QX, UK
\end{center}

\renewcommand{\LARGE}{\Large}
\renewcommand{\Huge}{\Large}

\vspace{.5in}

\abstract{
In minimal SUSY-GUT models with $M_{SUSY}\alt 1$ TeV, the renormalization
group equations
have a solution dominated by the infrared fixed point of the top Yukawa
coupling. This fixed point predicts $m_t\simeq (200\; {\rm GeV})\sin \beta $;
combined with the LEP results it excludes $m_t\alt 130$ GeV. For $m_t$
in the range 130--160 GeV, it predicts that the lightest scalar $h$ has mass
60--85 GeV (detectable at LEP\,II).
At SSC/LHC, each of the five scalars $h$, $H$, $A$, $H^{\pm }$ may be
detectable, but not all of them together; in one parameter region none would
be detectable.}

\vspace{.5in}

For a large top-quark mass $m_t>M_W$, the corresponding Yukawa coupling
$\lambda_t$ is plausibly large at the GUT scale $M_G$, in which case
$\lambda_t$ evolves rapidly toward an infrared fixed point at low mass
scales\cite{pendleton,ramond,dhr,bbo,knowles,pokorski}. The evolution of the
top quark Yukawa
coupling $\lambda _t$ is governed by
the one-loop renormalization group equation
\begin{equation}
{{d\lambda _t}\over {dt}}={{\lambda _t}\over {16\pi ^2}}\left [
-\sum c_ig_i^2+6\lambda _t^2+\lambda _b^2\right ]\;,\label{lambdat}
\end{equation}
with $c_1=13/15$, $c_2=3$, $c_3=16/3$; the couplings evolve toward a fixed
point close to
where the quantity in square brackets in Eq.~(\ref{lambdat}) vanishes.
Then the known gauge couplings determine the running
mass $m_t(m_t)=\lambda_t(m_t)v\sin\beta/\sqrt2$ and hence the pole mass
$m_t({\rm pole})=m_t(m_t)\left[1+{4\over3\pi}\alpha_s(m_t)\right]$; two-loop
evaluations\cite{bbo} give
\begin{equation}
m_t({\rm pole})\simeq(200\gev)\sin\beta \,,  \label{mtpole}
\end{equation}
where $\tan\beta=v_2/v_1$ is the usual ratio of two Higgs vevs. If
$\lambda_t(M_G)$ is below the fixed point, its convergence to the fixed point
is more gradual and Eq.~(\ref{mtpole}) does not necessarily apply. But in
practice large $\lambda_t(M_G)$ is favored in many SUSY-GUT solutions;
large $\lambda _t(M_G)$ facilitates
 $\lambda_b(M_G)=\lambda_\tau(M_G)$
Yukawa unification\cite{CEG} and
allows intricate relationships between fermion
masses and mixings.
It is
therefore interesting to pursue the phenomenological implication of
Eq.~(\ref{mtpole})~\cite{bbop}.

Figure 1 shows how $\lambda_t(M_G)$ and $\lambda_b(M_G)$ values
relate to $m_t(\rm pole)$ and
$\tan\beta$ in SUSY-GUT solutions with Yukawa unification; the lower (upper)
shaded
branches contains the $m_t\ (m_b)$ fixed-point solutions. There is a small
region at the upper right where both fixed point solutions are simultaneously
satisfied. Figure~2 shows that
the $m_t$ fixed-point behavior is insensitive to GUT threshold corrections in
the $\lambda_b/\lambda_\tau$ ratio, at least for threshold corrections
$\alt 10\%$. The
sensitivity of the fixed point to threshold corrections is decreased
for larger values of $\alpha _s(M_Z)$ where the solutions tend to have a
stronger fixed point character, as indicated by Eq.~(\ref{lambdat}). The
perturbative limits of the Yukawa couplings near their Landau poles are
shown in Fig.~2(a) as the dashed lines $\lambda _t^G = 3.3$ and
$\lambda _b^G = 3.1$.

\begin{center}
\parbox[c]{2.875in}{
\epsfxsize=2.875in
\hspace{0in}
\epsffile{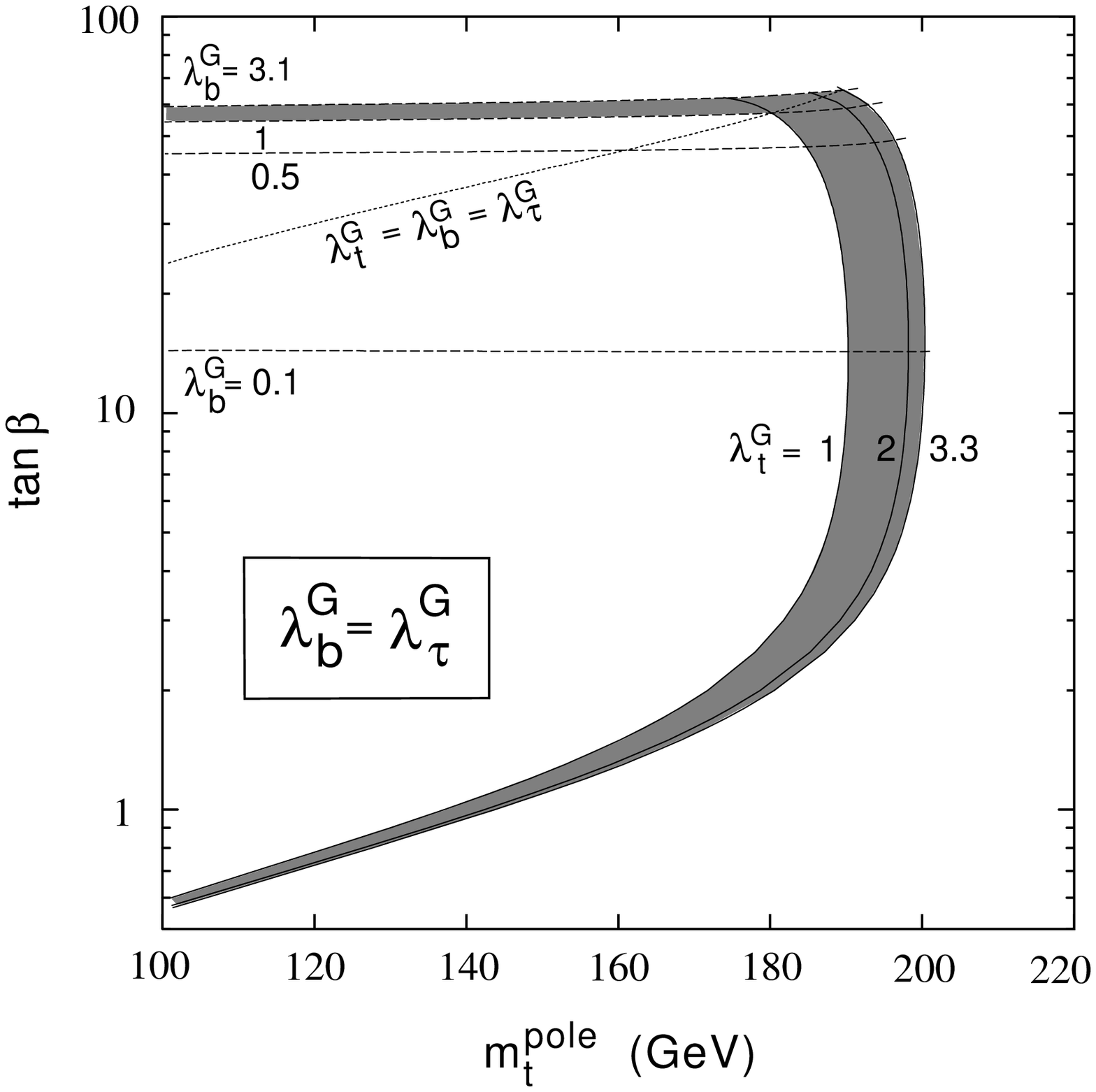}
}
\parbox[c]{2.875in}{\small Fig.~1: Contours of constant Yukawa couplings
$\lambda _i^G=\lambda _i(M_G^{})$
at the GUT scale in the ($m_t^{\rm pole}, \tan \beta$) plane,
obtained from solutions to the RGE with
$\lambda _{\tau}^G=\lambda _b^G$ unification imposed.
The GUT scale Yukawa coupling contours are close together for
large $\lambda ^G$. The
fixed points describe the values of the Yukawa couplings at
the electroweak scale for $\lambda _t^G\agt 1$ and
$\lambda _b^G\agt 1$.}

\end{center}

The minimal SUSY Higgs spectrum contains two CP-even scalars $h$ and $H$
($m_h<m_H$), a CP-odd pseudoscalar $A$ and two charged scalars $H^\pm$. At tree
level their properties are controlled by two parameters $m_A,\tan\beta$.
Radiative corrections depend principally on $m_t$ (constrained now by
Eq.~\ref{mtpole}) and logarithmically on $m_{\tilde t}$, that we here set at
$m_{\tilde t}=1$~TeV. Assuming $m_t\alt160$~GeV, Eq.~(\ref{mtpole}) constrains
$\tan\beta$ to values near 1, where $h$ is relatively light (recall the
tree-level relation $m_h<M_Z|\cos 2\beta |$) and the couplings of $h$ are close
to those of a Standard Model Higgs boson.
LEP Higgs searches\cite{lep,lepsm} exclude a region of the
$(m_A,\tan\beta)$ plane shown in Fig.~3(a): this translates to forbidden
regions in $(m_h,\tan\beta)$ in Fig.~3(b). We see that the fixed-point
condition
predicts $m_t\agt 130$~GeV, $m_h\agt60$~GeV, $m_A\agt70$~GeV; correspondingly
$m_{H^\pm}\agt105$~GeV,  $m_H\agt140$~GeV. If in fact $m_t\alt160$~GeV, then
$m_h\alt85$~GeV as shown in Fig.~4, and $h$ will be discoverable at LEP\,II
(but none of the
other Higgs bosons will). The discovery limits at SSC/LHC (taken here from
Ref.~\cite{bcps}) are also
shown in Fig.~4; we see that each of the five Higgs bosons might be
discoverable there, but not all at once, and possibly none of them at
all.

\begin{center}
\epsfxsize=5.75in
\hspace{0in}
\epsffile{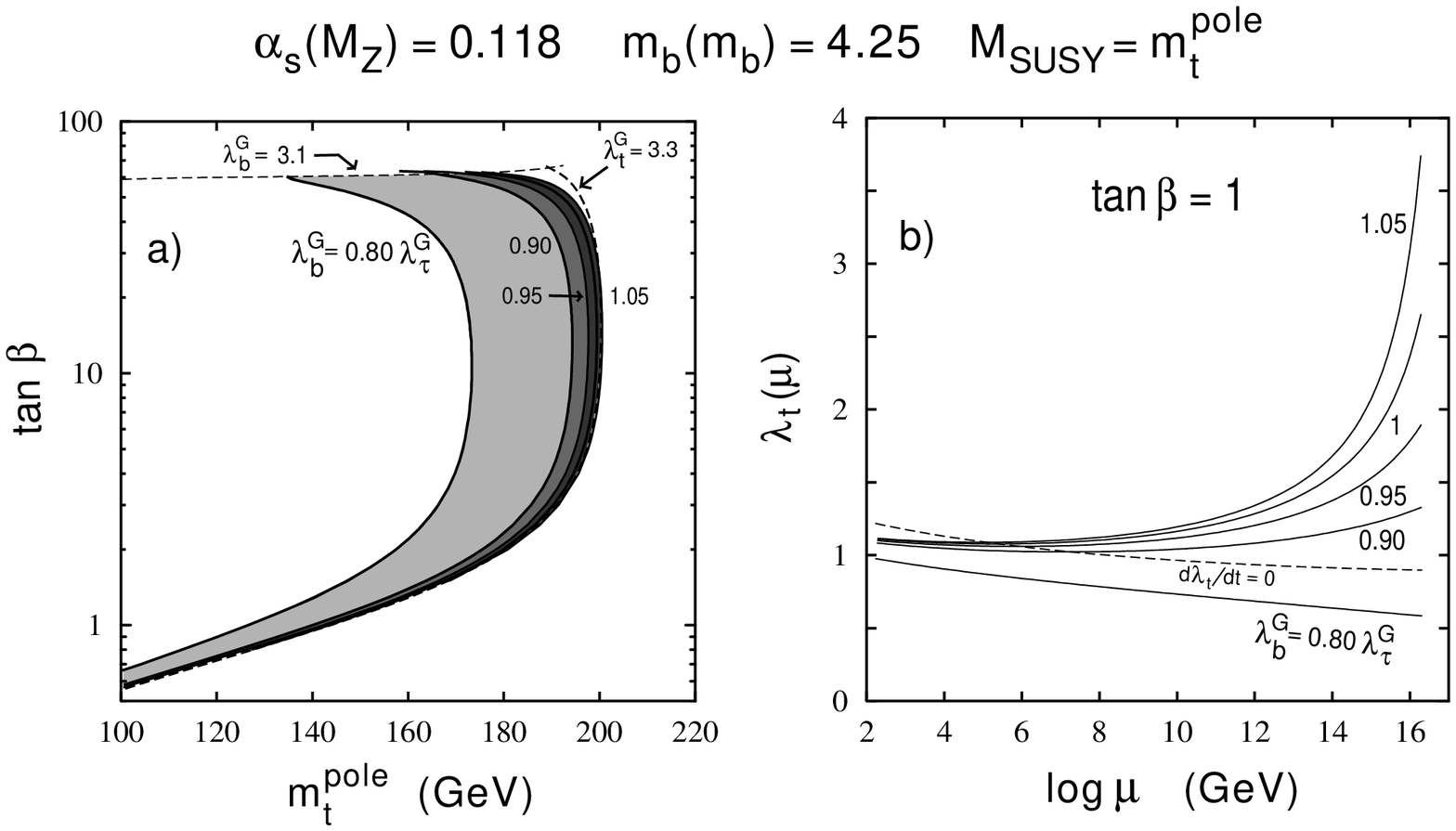}

\parbox{5.75in}{\small Fig.~2: RGE results for
$\alpha _s(M_Z^{})=0.118$ with the boundary condition $m_b(m_b)=4.25$
GeV.
(a) GUT threshold corrections to Yukawa coupling unification.
The solutions strongly exhibit
a fixed point nature, for threshold
corrections $\alt 10\%$.
Taking a larger supersymmetric threshold $M_{SUSY}^{}$
or increasing $\alpha _s(M_Z)$
moves the curves to the right, so that the fixed point condition becomes
stronger.
(b)~Evolution of the top quark Yukawa coupling for
$\tan \beta =1$. The dashed line indicates
${{d\lambda _t}\over {dt}}=0$ which gives an
approximation to the electroweak scale value of $m_t$ with accuracy of
order~10\%.}

\end{center}

\begin{center}

\epsfxsize=5.5in
\hspace{0in}
\epsffile{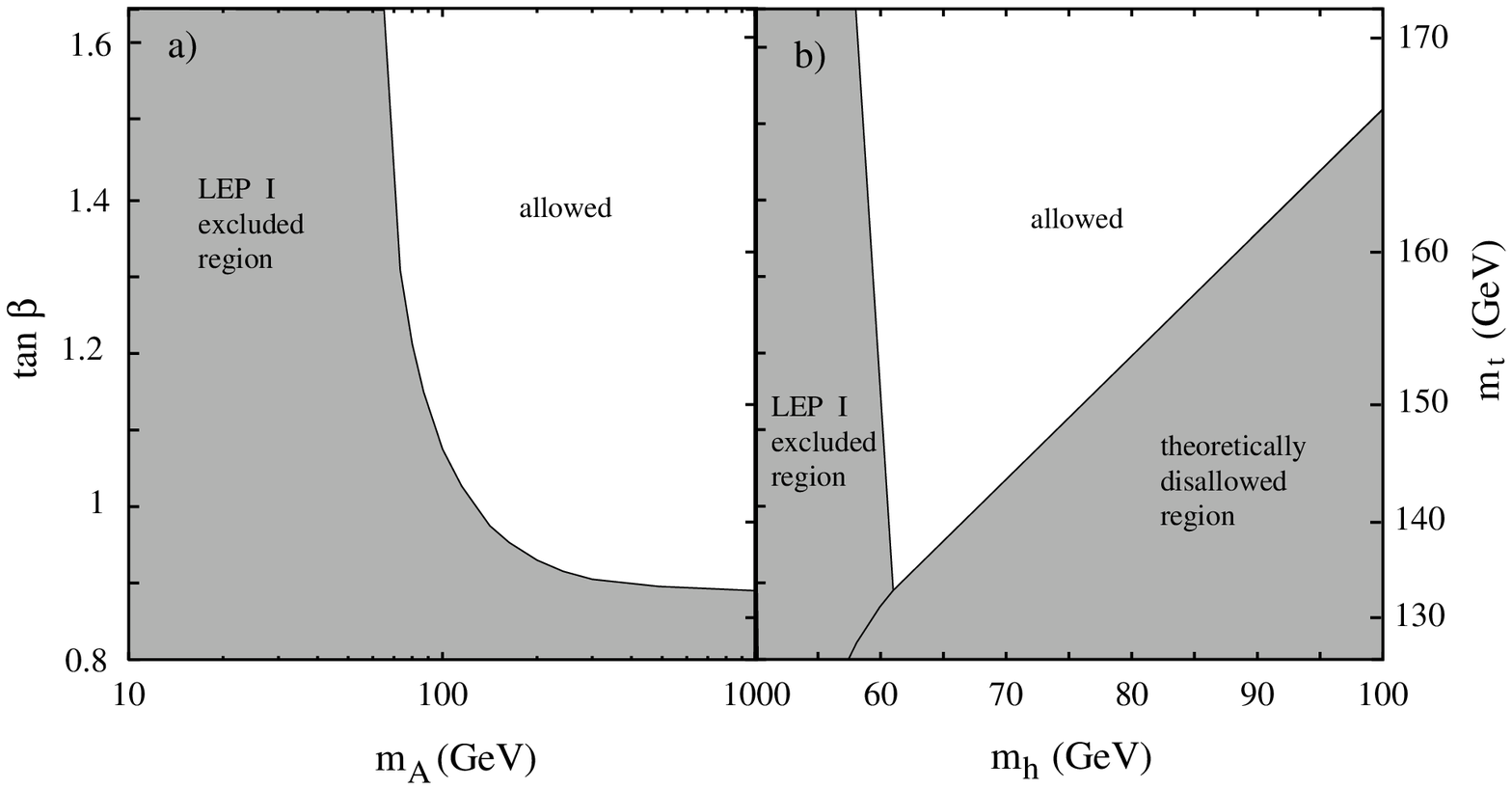}

\parbox{5.5in}{\small Fig.~3: $m_t$ fixed-point solution regions allowed by the
LEP\,I data: (a)~in the $(m_A, \tan \beta )$ plane, (b)~in the $(m_h, \tan
\beta )$ plane. The top quark masses are $m_t({\rm pole})$, correlated to
   $\tan \beta $ by Eq.~(\ref{mtpole}).}

\bigskip

\parbox[c]{3in}{
\epsfxsize=3in
\hspace{0in}
\epsffile{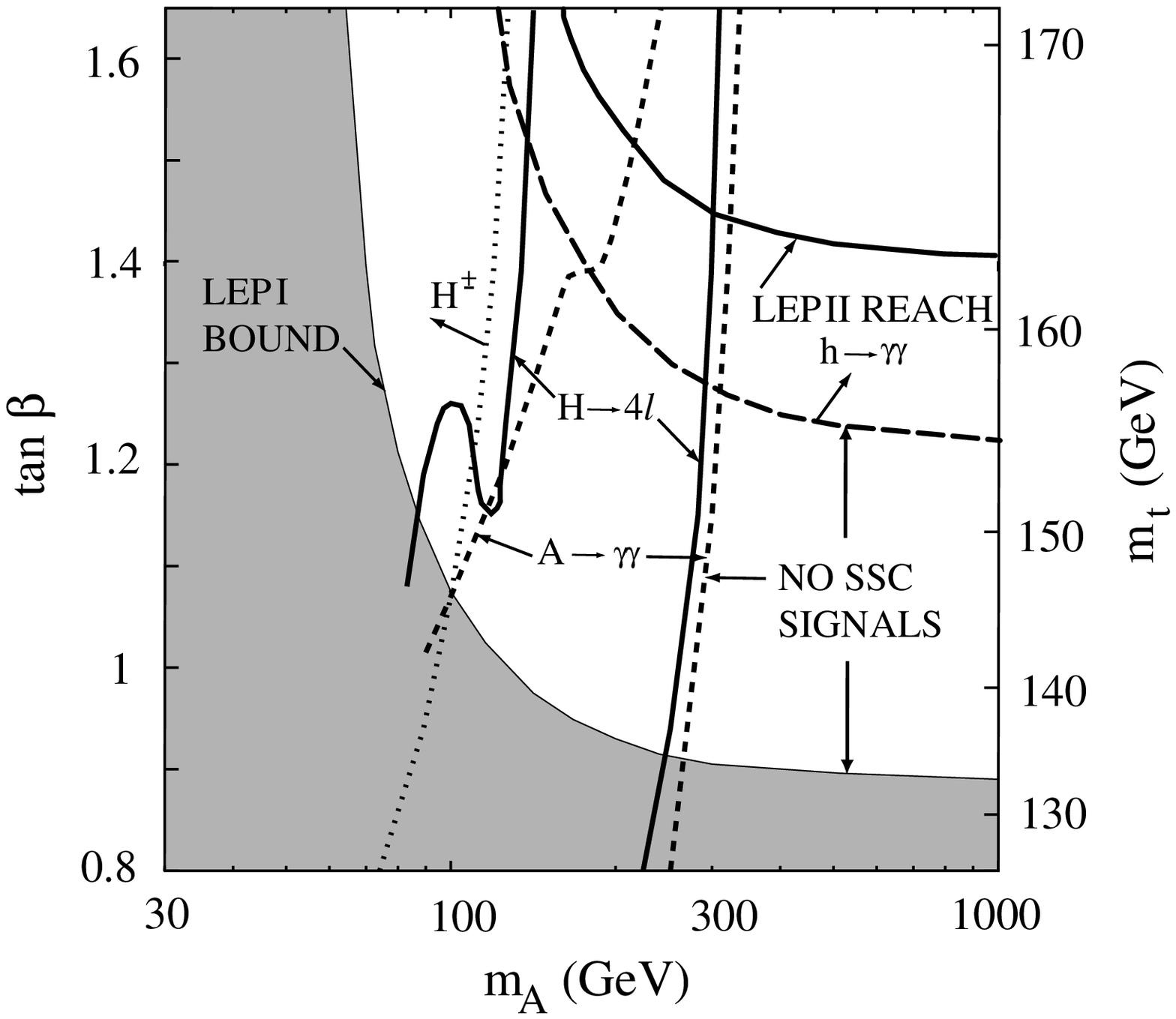}
}\hfill
\parbox[c]{2.75in}{\small Fig.~4: SSC/LHC signal detectability regions,
compared with the LEP\,I allowed region of $m_t$ fixed-point solutions from
Fig.~3(a) and the
probable reach of LEP\,II. The top quark masses are $m_t({\rm pole})$.}

\end{center}

{\small This work was supported in part by DE-AC02-76ER00881 and TNLRC
RGFY9273.}


\vspace*{-0.25in}


\begin{thebibliography}{0}


\bibitem{pendleton} B.~Pendleton and G.~G.~Ross,
   Phys.\ Lett.\ {\bf B98}, 291 (1981); C.~T.~Hill,
   Phys.\ Rev.\ {\bf D24}, 691 (1981).


\bibitem{ramond} H.~Arason, et al., Phys.\ Rev.\ Lett.\
   {\bf 67}, 2933 (1991); Phys.\ Rev.\ {\bf D47}, 232 (1993).

\bibitem{dhr} S.~Dimopoulos, L.~J.~Hall and S.~Raby, Phys.\ Rev.\ Lett.
   {\bf 68}, 1984 (1992); Phys.\ Rev.\ {\bf D45}, 4192 (1992);
   G.~F.~Giudice, Mod.\ Phys.\ Lett.\ {\bf A7}, 2429 (1992).

\bibitem{bbo} V.~Barger, M.~S.~Berger, and P.~Ohmann, Phys.\ Rev.\
   {\bf D47}, 1093 (1993); V.~Barger, M.~S.~Berger, T.~Han and M.~Zralek,
   Phys.\ Rev.\ Lett.\ {\bf 68}, 3394 (1992).

\bibitem{knowles} C.~D.~Froggatt, I.~G.~Knowles and R.~G.~Moorhouse,
   Phys.\ Lett.\ {\bf B249}, 273 (1990);
   Phys.\ Lett.\ {\bf B298}, 356 (1993).

\bibitem{pokorski} M.~Carena, S.~Pokorski, and C.~E.~M.~Wagner, Munich
   preprint MPI-Ph/93-10.

\bibitem{CEG} M.~Chanowitz, J.~Ellis and M.~Gaillard, Nucl.\ Phys.\ {\bf B128},
506 (1977).

\bibitem{bbop} For details see V.~Barger et al., University of
Wisconsin-Madison preprint MAD/PH/755.

\bibitem{lep} ALEPH Collaboration: D.~Decamp et al, Phys.\ Lett.\
   {\bf B246}, 623, (1990), {\bf B265}, 475 (1991); DELPHI
   Collaboration: P.~Abreu et al, ibid {\bf B245}, 276 (1990), Nucl.\
   Phys.\ {\bf B373}, 3 (1992); L3 Collaboration: B.~Adeva et al,
   Phys.\ Lett.\ {\bf B294}, 457 (1992);  OPAL Collaboration:
   M.~Z.~Akrawy et al, Z.\ Phys.\ {\bf C49}, 1 (1991).

\bibitem{lepsm} T.~Mori, report to Dallas Conference 1992; E.~Gross
   and P.~Yepes, CERN-PPE/92-153.

\bibitem{bcps} V.~Barger et al., Phys.\ Rev.\ {\bf D46}, 4914 (1992).

\end{thebibliography}
\end{document}